\documentclass[cits]{PoS}

\title{The elusive radio loud Seyfert 2 galaxy NGC 2110}

\ShortTitle{The elusive radio loud Seyfert 2 galaxy NGC 2110}

\author{\speaker{Volker Beckmann}\\%
        Fran\c{c}ois Arago Centre, APC, Universit\'e Paris Diderot, CNRS/IN2P3, CEA/DSM, Observatoire de Paris, 13 rue Watt, 75205 Paris Cedex 13, France\\
        E-mail: \email{beckmann@apc.univ-paris7.fr}}

\author{Olivier Do Cao\\
        Laboratoire AIM Paris-Saclay, CEA/Irfu Universit\'e Paris-Diderot CNRS/INSU, 91191 Gif-sur-Yvette, France\\
        E-mail: \email{olivier.do-cao@cea.fr}}

\abstract{The AGN NGC~2110 presents a peculiar case among the Seyfert~2 galaxies, as it displays also features of radio-loud objects and is classified as FR-I radio galaxy. Here we analyse simultaneous {\it INTEGRAL} and {\it Swift} data taken in 2008 and 2009. We reconstruct the spectral energy distribution in order to provide further insight. The combined X-ray spectrum is well represented by an absorbed cut-off power law model plus soft excess. Combining all available data, the spectrum appears flat ($\Gamma = 1.25 \pm 0.04$) with the high-energy cut-off being at $82 \pm 9 \rm \, keV$. The absorption is moderate ($N_{\rm H} = 4 \times 10^{22} \rm \, cm^{-2}$), the iron K$\alpha$ line is weak ($EW = 114 \rm \, eV$), and no reflection component is detected in the {\it INTEGRAL} spectrum. The data indicate that the X-ray spectrum is moderately variable both in flux and spectral shape. The 2008 spectrum is slightly steeper ($\Gamma = 1.5$, $E_C = 90 \rm \, keV$) with the source being brighter, and flatter in 2009 ($\Gamma = 1.4$, $E_C = 120 \, \rm keV$) in the lower flux state. The spectral energy distribution gives a bolometric luminosity of $L_{bol} \simeq 2 \times 10^{44} \rm \, erg \, s^{-1}$. NGC~2110 appears to be a borderline object between radio loud narrow line Seyfert~1 and radio quiet Seyfert~2. Its spectral energy distribution might indeed be dominated by non-thermal emission arising from the jet.
}

\FullConference{8th INTEGRAL Workshop "The Restless Gamma-ray Universe"\\
                 September 27-30 2010\\
                 Dublin Castle, Dublin, Ireland}

\begin{document}

\section{Introduction}

NGC 2110 is a nearby ($z=0.0078$, $D=34.5$ Mpc) S0 galaxy first discovered in the X-rays in the late 1970s. 
It was first classified as a Seyfert~1 galaxy particularly because of its X-ray luminosity and its timescale variability. 
But e.g. Malaguti et al. \cite{malaguti99} suggested it to be classified as a Narrow Emission Line Galaxy (NELG), a transition between Seyfert~1 and Seyfert~2, because of its moderately flat ($\Gamma \simeq 1.7$) powerlaw spectrum in the X-rays and narrow emission lines in the optical spectrum. On the other hand, the total extent of the radio structure is 
$4''$ (830 pc), similar to a radio loud galaxy \cite{ulvestad83}, the source is of FR-I type with the radio jet being clearly visible. The radio flux is $\sim 300 \rm \, mJy$ at $1.4 \rm \, GHz$, as measured in the NVSS, which gives a radio loudness $\log (f_R / f_V) = 1.2$ when applying an aperture of $22.5"$ for the optical measurement of the core \cite{McAlary83}.

This is supported by the fact that, unlike an average Seyfert galaxy, NGC~2110 contains a flat spectrum nuclear radio component. 
Thus on one hand NGC~2110 presents simultaneously characteristics of a radio quiet (high X-ray luminosity, emission lines) and a radio loud (radio structure, spectral energy distribution overview) galaxy. On the other hand, even inside the radio quiet class, this galaxy shows both type~1 (e.g. flat continuum) and type~2 (e.g. narrow emission lines only) features. Those elements suggest that NGC~2110 could be a peculiar galaxy, inconsistent with the unified models or belonging to a transition class between radio quiet and radio loud galaxies.

 The aim of this work is to test the unified model through the investigation of this peculiarity in the X-rays, providing a direct view on the central engine emission mostly free of absorption effects. The combined data, recently acquired by the {\it INTEGRAL} and {\it Swift} satellites, give new insight into the nature of NGC~2110. We will first extract and model the X-ray spectrum, and then add these results into the spectral energy distribution in order to estimate the bolometric luminosity. 

\section{Data analysis}

In this study we use the {\it INTEGRAL} data taken in two observation campaigns in August/September 2008 and in October 2009. During these observations NGC 2110 was observed with IBIS/ISGRI for 158~ks and 190~ks, respectively. Because the observations were performed in hexagonal dithering mode and for some observations both JEM-X instruments were switched on, the total JEM-X exposure exceeds the IBIS/ISGRI one.  Together with the {\it INTEGRAL} campaign, short {\it Swift} snap-shot observations were performed in 2008 and 2009, with 2.2 ks  and 3.6 ks, respectively. Details about the exposure times are listed in Table~\ref{fitsummary}. 
We selected the data up to an off-axis angle of $10^\circ$ for IBIS/ISGRI and SPI, and within $3^\circ$ for JEM-X. {\it INTEGRAL} data reduction was performed using the Offline Scientific Analysis (OSA) package version 9.0. The {\it Swift}/XRT data were analysed using the mission specific software included in HEAsoft version 6.7 and using the latest calibration files.
We combined IBIS/ISGRI, JEM-X, and Swift/XRT data in order to achieve a broad band coverage from 0.1 to 300 keV. Spectra were extracted for each observation campaign, and a combined spectrum was derived including all available {\it INTEGRAL} and {\it Swift} data. In order to account for flux variability, an intercalibration factor has been applied. With respect to {\it Swift}/XRT, the IBIS/ISGRI derived flux appears to be 5\% higher, while JEM-X data require an intercalibration factor of 1.7.

As spectral model we applied first an absorbed power law which gave a bad representation of the data in all cases ($\chi^2_\nu > 1.7$). Adding a high-energy cut-off improved the fit significantly with a confidence level greater than 99.9\% according to an F-test. For the different data sets the cut-off is in the range 80 keV to 120 keV, with a photon index of the underlying continuum between $\Gamma = 1.25$ and $\Gamma = 1.35$, with $\chi^2_\nu > 1.25$ in all cases. The largest contribution to the residuals is originating from an excess in soft X-rays ($E < 2 \rm \, keV$) in the {\it Swift}/XRT spectra. In order to account for the excess, we applied a black body and a Raymond-Smith model \cite{Raymond77}. Both models gave a significantly improved fit result ($\chi^2_\nu \simeq 1$). Concerning the Raymond-Smith model, the abundance of the plasma could not be constrained and was fixed to the solar value. The fit results for the best fit model, i.e. an absorbed cut-off power law model plus soft excess, are summarized in Table~\ref{fitsummary}, using the black body model to account for the soft excess. In the case of the 2009 spectrum the energy of the soft excess is unconstrained and has been fixed to $kT_{bb} = 1 \rm \, keV$.
\begin{table}
\caption{Exposure times and spectral fit results for the absorbed cut-off power law model plus soft excess.}
\label{fitsummary}
\centering 
\begin{tabular}{c c c c c}
\hline\hline                
parameter & 2008 & 2009 & all data \\ 
\hline
ISGRI exposure [ks] & 158 & 190 & 347\\
JEMX-1 exposure [ks] &168 & 155 & 303\\
JEMX-2 exposure [ks] & 110 &  45 & 155\\
{\it Swift}/XRT exposure [ks] &    2.2& 3.6& 25.6\\

$\chi^2_\nu$ (dof) & 1.00 (41) & 1.28 (48) & 1.03 (204)\\
photon index $\Gamma$           & $1.35 \pm 0.09$ & $1.47 \pm 0.06$ & $1.25 \pm 0.04$\\
$E_C \rm \, [keV]$ & $87 \pm 23$ & $117 \pm 23$ & $82 \pm 9$\\
$N_{\rm H} \, \rm [10^{22} \, cm^{-2}]$ & $5.9 \pm 1.1$ & $4.5 \pm 0.1$ & $3.7 \pm 0.2$\\
$kT_{bb} \rm \, [keV]$ & $0.8 \pm 0.6$ & $1$ & $0.7 \pm 0.3$\\ 
$f_{20 - 100 \rm \, keV} [10^{-11} \rm \, erg \, cm^{-2} \, s^{-1}]$ & 17     & 31  & 24 \\
\hline
\end{tabular}
\end{table}

The iron line is detected with more than $2\sigma$ significance only in the combined data set. We fit this line by a Gaussian profile with the frozen parameters $E = 6.35 \rm \, keV$ and $\sigma = 0.08 \rm keV$ \cite{hayashi96}, resulting in an equivalent width of $EW = 114 {+17 \atop -41} \rm \, eV$. A more complex model, like a Comptonisation model, did not improve the fit further.
  


\section{Spectral energy distribution}

We reconstruct the spectral energy distribution (SED) of NGC~2110 in order to determine the bolometric luminosity and to get an overall view on the main emission components. 

In order to achieve a broad coverage in energy, we added data available from the literature which we retrieved from the NASA Extragalactic Database (NED). For the X-rays we use the spectrum described in the previous section. 
We then added the absorption corrected {\it INTEGRAL}/OMC measurement, assuming for the extinction $A_V=1.04 \rm \, mag$ \cite{elvis89}. 
%
%
From $m_V = 12.7 \rm \, mag$, we obtain $\nu F_{\nu} = 4.5 \times 10^{13} \rm \, Jy \, Hz$ at a frequency $\nu \simeq 5.5 \times 10^{14} \rm \, Hz$.

\begin{figure}
   \centering
   \includegraphics[width=10cm]{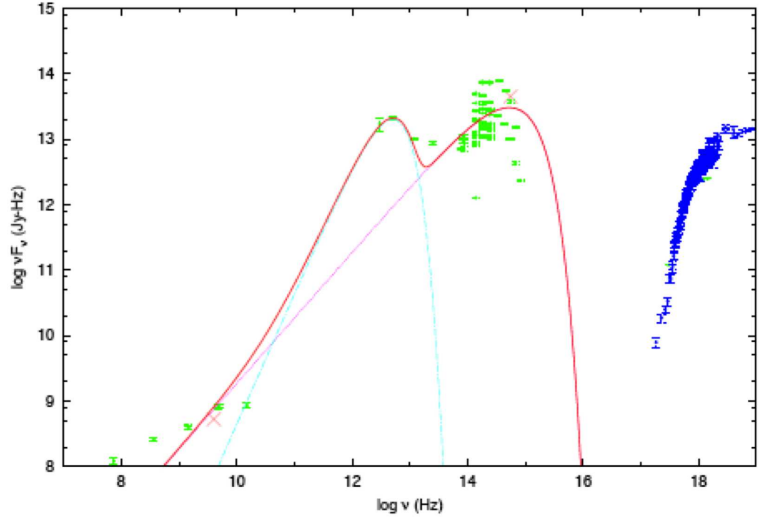}
   \caption{Spectral energy distribution of NGC~2110. Green and blue datapoints are from the NED and X-rays of this work respectively. The red crosses are the assumed values for the fit of the synchrotron branch (red solid line) which is the sum of the pure synchrotron radiation (mauve dotted line) and thermal dust emission (cyan dotted-dashed line).}
              \label{fit_sed}
    \end{figure}

As our data do not cover the whole SED simultaneously, our result will provide only a rough estimate of $L_{bol}$. In our modelisation, the SED can be divided into two main parts. The first one covers the radio to optical range, called the synchrotron branch because of the assumed dominance of the synchrotron radiation. The second one covers the X-rays, called Inverse Compton branch. They are represented by green and blue points in Figure~\ref{fit_sed}, respectively. 

We assume the energy output to be created by synchrotron radiation from a relativistic electron population
situated in the AGN core. The electrons are assumed to be spatially distributed as an homogeneous sphere of radius $R=1.4 \times 10^{-4} \rm \, pc$, based on the variability timescale of $t_{var}=4 \rm \, h$ \cite{hayashi96}. Following \cite{ballo02,salas95} we apply $\gamma_1=1$, $\alpha=\pi/2$, $p=1.0$, and $\gamma_2=5 \times 10^4$, leaving the magnetic field $B$ and the electron density $K$ free to vary.

NGC 2110 shows large variations in the optical and we used for the model the OMC measurement to fix the peak of the synchrotron emission. 
The model we apply does not predict a flat radio continuum. We therefore take the average value of the five datapoints as one single measurement. 
A secondary peak in the far-IR ($\nu \simeq 5 \times 10^{12}$ Hz) is assumed to be the thermal emission from the dusty torus at $T_{dust}=85 \rm \, K$. 
We used numerical integration 
 to compute the theoretical spectrum. The best fit gives a magnetic field with $B=0.008 \rm \, G$ and an electron density $K=6 \times 10^9 \rm \, cm^{-3}$.

From the pure synchrotron contribution, we derive the emitted luminosity $L_{sync} = 1.2 \times 10^{44} \rm \,erg \, s^{-1}$. The X-ray contribution to the bolometric luminosity is given by the spectrum presented here: $L_{IC}=6.3 \times 10^{43} \rm \, erg \, s^{-1}$, consistent with earlier {\it INTEGRAL} results \cite{Beckmann09}. As {\it INTEGRAL} and {\it Swift} do not cover the entire X-ray/$\gamma$-ray range, we extrapolated the model from 0.1 keV to 10 MeV. Combining these results we estimate the bolometric luminosity to be $L_{bol} \simeq 2 \times 10^{44} ~\rm \, erg \, s^{-1}$.

\section{Discussion}

An absorbed cut-off powerlaw model ($E_C \simeq 80 \rm \, keV$, $\Gamma \simeq 1.3$) plus soft excess ($kT_{bb} \simeq 0.8 \rm \, keV$) is sufficient to represent the X-ray data on NGC~2110 provided by {\it INTEGRAL} and {\it Swift}. 
{\it BeppoSAX} data showed a steeper spectrum and gave only a lower limit for the cut-off energy ($E_C > 70 \rm \, keV$, $\Gamma = 1.74 {+0.20 \atop -0.13}$, \cite{Dadina07}). {\it Suzaku} spectra in the $0.5 - 50 \rm \, keV$ energy range were fit by a simple power law model with $\Gamma = 1.59 \pm 0.01$ \cite{Miyazawa09}. Also combined  {\it Chandra} and {\it XMM-Newton} data can be represented by a simple power law model with $\Gamma = 1.74 \pm 0.05$ \cite{Evans07}. When applying a simple power law to the data presented here we get a photon index of $\Gamma = 1.6$, consistent with these results. The weak iron line with $EW = 114 \rm \, eV$ had been detected before with equivalent widths in the range of $100 - 200 \rm \, eV$ \cite{malaguti99, Dadina07, Evans07}.
On average, Seyfert galaxies have $\Gamma \sim 1.5$ and $E_c \sim 100 \rm \, keV$ \cite{nandra94}. NGC~2110 is intrinsically slightly flatter  with $\Gamma=1.3$ and $E_C \simeq 80 \rm \, keV$. 

Concerning the soft excess, we have only tested the blackbody and the Raymond-Smith models. The former is the simplest model while the latter is the commonly accepted one for a two-phase disc \cite{haardt93}. Again, our data do not permit us to favour one model over the other but our measurement ($k_BT \simeq 1.4 \rm \, keV$) is consistent with previous work within the error range. 
However, the soft excess, if modeled by the blackbody tail produced by the $\alpha$-disc, is physically not satisfying and we thus prefer the interpretation of the soft excess being represented by the Raymond-Smith model, even if other models cannot be ruled out. 

In order to determine the spectral energy distribution from the radio to optical, we estimated the source to be a single relativistic electron population homogeneously distributed in a sphere of radius $R=1.4 \times 10^{-4} \rm \, pc$ with a density of $K=6.0 \times 10^9 \rm \, cm^{-3}$, and a magnetic field of $B=0.008 \rm \, G$. These values are different compared to typical magnetic field $B \sim 10^{-6} \rm \, G$ \cite{parker79} and density $n \sim 10^{15} \rm cm^{-3} $ \cite{netzer06}, probably because of the oversimplifications used in our model. However, the derived luminosities are consistent with previous works and are typical of Seyfert galaxies \cite{Beckmann09,vasudevan10}. From this point of view, NGC~2110 does not differ from other Seyfert galaxies. Nevertheless, this source displays a strong synchrotron component, visible at least in the radio domain. Typical radio loud galaxy luminosities are though higher by at least one order of magnitude \cite{elvis94}. The scatter of the optical measurements indicates that the bolometric luminosity of NGC~2110 is highly variable.

NGC~2110 shows evidence for peculiarities. The inner structure is highly complex, as seen in the soft X-rays \cite{Evans07} and in the radio domain, where NGC~2110 exhibits not only a complex and extended radio structure, but also strong nuclear variability \cite{Mundell09}. The overall SED might indeed be dominated by non-thermal emission, especially in the radio and X-ray domain. With $\alpha_{RX} = 0.7$ between $1.4 \rm \, GHz$ and the X-ray band, this source has a similar X-ray to radio ratio as X-ray detected BL Lac objects which have $\langle \alpha_{RX} \rangle = 0.6$, and also $\alpha_{RO} = 0.2$ and $\alpha_{OX} = 1.7$ are consistent with a BL Lac type core \cite{Beckmann03}.  NGC~2110 presents a case similar to that of Cen~A, in which also a strong FR-I together with a Seyfert~2 are observed and the non-thermal component seems to dominate \cite{Beckmann11}. 
We suggest that NGC~2110 is a member of a transition class between radio loud and radio quiet galaxies because of its radio features with remaining Seyfert properties. An alternative scenario would be that the unified picture in which Seyfert type~1 and 2 galaxies belong to the same class of galaxies viewed under different angles, may be in fact more complex. Future investigations should also consider flux variability in the interpretation of the data. It would be even more valuable to obtain simultaneous data in order to avoid these effects. Finally, measurements in the UV or in the sub-mm domain are required to confirm the presence of strong broad-band synchrotron radiation.

\end{document}